# An Overview of Extremely Large Telescopes Projects

*R. G. Carlberg*

Department of Astronomy and Astrophysics,
University of Toronto, Toronto M5S 3H8, Canada
email: carlberg@astro.utoronto.ca

**Abstract.** IAU Symposium 232 allows a snapshot of ELTs at a stage when design work in several critical mass projects has been seriously underway for two to three years. The status and some of the main initial design choices are reviewed for the North American Giant Magellan Telescope (GMT) and the Thirty Meter Telescope (TMT) projects and the European Euro50 and the Overwhelmingly Large (OWL) projects. All the projects are drawing from the same "basket" of science requirements, although each project has somewhat different ambitions. The role of the project offices in creating the balance between project scope, timeline and cost, the "iron triangle" of project management, is emphasized with the OWL project providing a striking demonstration at this meeting. There is a reasonable case that the very broad range of science would be most efficiently undertaken on several complementary telescopes.

## 1. Introduction: The Motivation for ELTs

The convergence of three key factors around the year 2000 initiated several ELT design projects which required the internal allocation of one to two million of effort within observatories. These factors were new science, new technology and new methods of construction that would lower cost.

Ground-based optical-infrared telescopes made a string of truly remarkable and profound discoveries in the 1990's which can largely be attributed to the arrival of the 8m class telescopes. The Lyman-break technique allowed the selection of galaxies at much higher redshifts, in what is clearly the "epoch of galaxy formation". Supernova cosmology came forth with the startling discovery that expansion of the universe was accelerating. And, telescopes ranging from quite modest apertures (with reduced time pressure on them as the 8m class telescopes became available) up to the largest began discovering extra-solar planets by the dozens. These discoveries remind us that astrophysics has plenty of "discovery space" left to be explored with yet-larger telescopes.

All of the ELT science cases are built on the physical characterization of a range of the most important questions combined with new capabilities which should allow unanticipated discoveries. The main elements, in brief, are:
- Physical characterization and formation mechanisms of the extra-solar planets,
- The internal mechanisms for the formation of stars and galaxies,
- The nature of the "dark sector", dark matter and dark energy,
- Complementing the capabilities of JWST and ALMA, plus the whole range of "multi-wavelength" astronomy.

To undertake this science requires a telescope of at least 20m capable of operating in the diffraction limit much of the time in order to have the necessary sensitivity and angular resolution.

Ever since Galileo's development of the telescope for astronomical observations some four hundred years ago the value of bigger telescopes has been apparent, in that the amount of light gathered goes up with the aperture diameter, D, in proportion to the aperture area. In the seeing-limited regime the photon statistics of the noise is proportional to the square root of the sky brightness and the aperture area. The resulting "light bucket" telescope has a science return increasing as $D^2$ per unit time. Given that historical telescope costs have risen as $D^{2.7}$ (Meinel, 1978) the arguments of the last fifty years for building bigger telescopes generally emphasized better sites (from few arc-second images to sub-arcsecond images) to boost the science return into a regime where the benefits merited the increased cost. The ongoing Adaptive Optics "revolution" which provides nearly diffraction limited imaging allows the science return per unit time to increase as $D^4$, since the majority of observations have noise dominated by the photon statistics of the sky background.

## 2. The Technical and Financial Feasibility of ELTs

The science return per unit time from an ELT depends on the aperture, D, the ability of the AO system to provide diffraction limited images, simply measured as the Strehl ratio, S, and the fraction of the year spent in diffraction limited and seeing limited observations, $f_D$ and $f_{AO}$, respectively. Then, to a first approximation the science return is, $f_D D^4 S^2 + f_{AO} D^2$ per unit time. All of D, S and the sum $f_D + f_{AO} \leq 1$ are very expensive in both construction and operating costs. The aperture, D, is unique in that it is essentially fixed for the lifetime of the telescope, whereas improvements can, in principle, be made in AO systems and operational efficiency, although in practice both those improvements can be limited or made more difficult by the telescope design. The outcome is the range of tradeoff decisions has increased relative to older telescopes.

The ELT designs emerged about thirty years ago, although at the time they were "light buckets" and did not proceed to construction for a number of reasons (see, for instance, http://www.noao.edu/image_gallery/html/im0250.html). In the last decade there have been two technical developments of now proven importance, adaptive optics and the concept of mirror segmentation, both of which are made to work using the ongoing revolution in computing power. Adaptive optics is now being implemented on 8m class telescopes with great success. Even conservative assumptions about further improvements in lasers guide stars, wave-front sensors, deformable mirrors and computing will allow the construction of 30m class AO systems that will provide similar or better improvement. Telescope primary mirror segmentation is an extremely important concept that allows the primary mirror to be built out of "factory produced" small mirrors which are then phased together to create a single large mirror. The resulting primary is far thinner, hence lighter and with better thermal and other control properties, than any scale-up of a monolithic 8m mirror. The weight saving translates directly into reduced requirements on the telescope structure and its costs. Very substantial opto-mechanical control through measurement of performance parameters and computed feedback to actuators is required to maintain the mirror at the desired figure.

The near simultaneous arrival of science problems beyond the grasp of existing telescopes and the engineering reality of functioning AO systems and segmented mirrors was realized in astronomical communities around the world in about the year 2000. On that basis several existing observatories undertook internal studies of the feasibility of telescopes ranging from 20m to 100m aperture (for instance the MAXAT work of AURA), which then developed into substantial projects, coming to the conclusion that these telescopes could be built at a cost of about 1/3 to 1 billion dollars, which is deemed to be within the capabilities of a number of institutional partnerships.

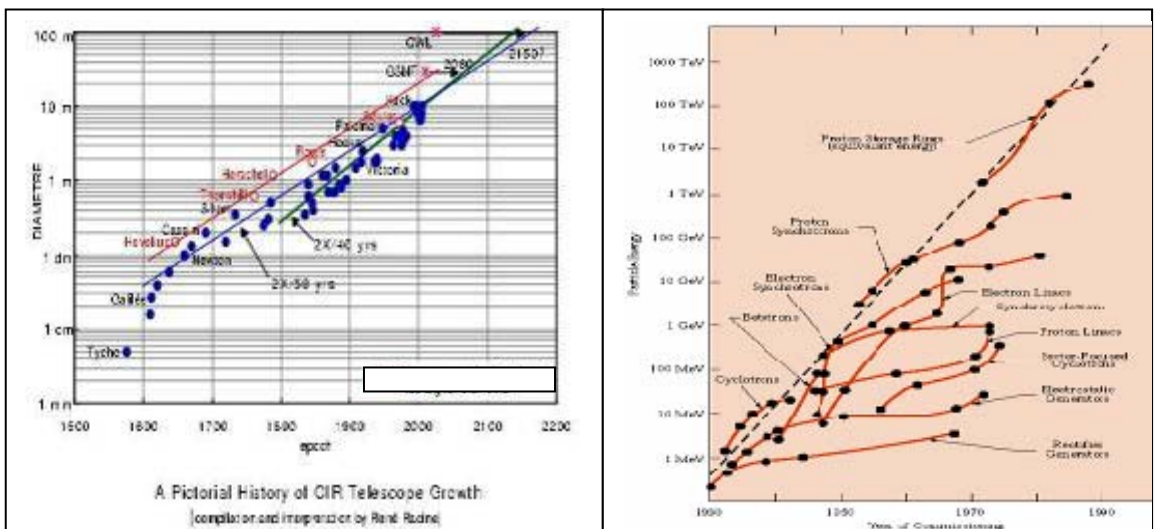

The growth of telescopes (left, courtesy R. Racine) and particle accelerators (left, from Panovsky's Beamline article). For four hundred years astronomers have been largely been increasing observation power by polishing ever larger glass mirrors doubling aperture in 40 years. Adaptive Optics and computing technology now available makes a faster growth of telescope power technically and financially feasible.

## 3. The Role of the Project Office

A telescope starts to become real when a project office is established. The job of the project office is to create the balance between the three sides of the "iron triangle" of project management: scope, cost and time. To be extremely successful requires a vigorous debate between scientists who defend the necessity of key science requirements, some of which may be beyond "ready-made" technical solutions, and the project staff, who are required to deliver an operational facility on time and within budget. Without intense engagement of both sides of this equation the resulting telescope will miss its goals in some way.

The following images over-simplify the relations between the various project elements. Individual astronomers looking forward to an exciting project see the telescope entirely as it performs to their own expectations. On the other hand the project manager sees a large and very diverse community of astronomers insistent on their disparate needs bearing down on him. Given this tendency to scientifically overload the plan, the Board is constantly concerned that the available resources of money and people will not be sufficient to make even the basics work. However, the 8m era has taught us that astronomers, engineers and managers have developed the structures and attitudes that allow the decisions to be made that lead to successful projects, from all viewpoints.

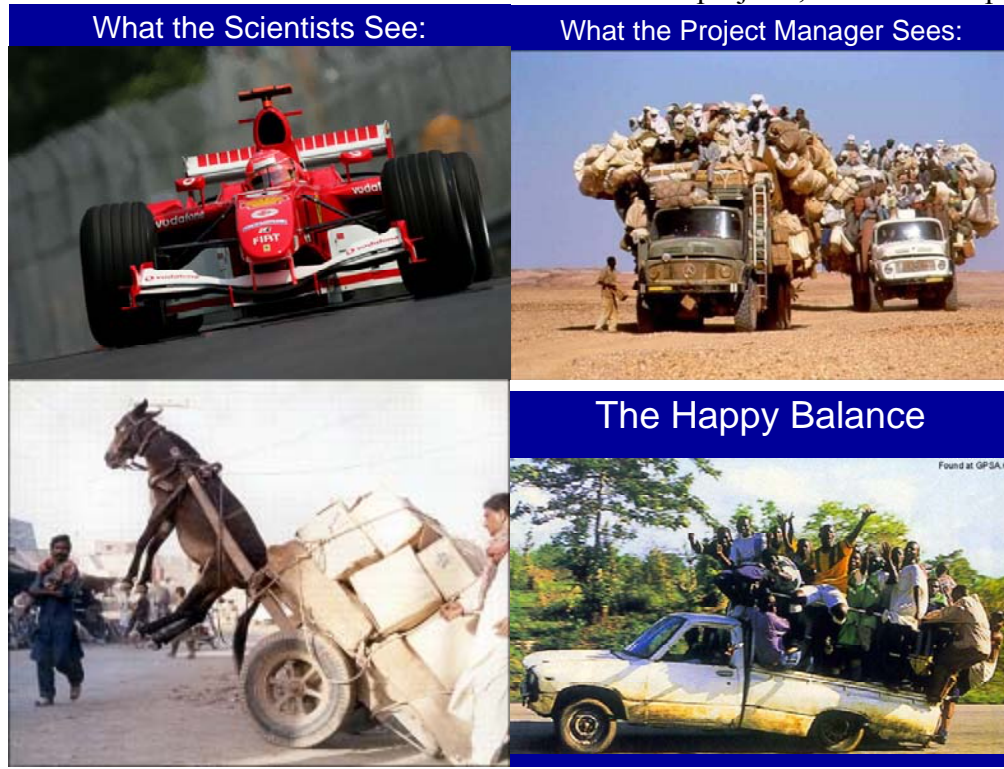

## 4. The Current Critical Mass ELT Projects

Once the basic technical and financial feasibility was established, the next step requires a significant commitment of resources plus the formation of a partnership with sufficient future resource prospects that a conceptual design can be presented to the partnership for construction and operation funds. The largest astronomy organizations in the world then become the focus for design partnerships and naturally fall into the North American public-private partnerships and the European collaborations based on combinations of national observatories, ESO and OPTICON. In both cases there are new funding arrangements being explored to ensure that the resources for these expensive ventures can be found.

In North America there are two project partnerships, the Giant Magellan Telescope and the Thirty Meter Telescope. Both involve AURA as a partner which proposes to seek funding from the National Science Foundation on behalf of astronomers through the USA. At the present time it is expected that this will lead to a competition between the two projects in about 2008 to be selected as the funding partner. In Europe there is the fairly long standing Euro 50 collaboration and the ESO sponsored Overwhelming Large telescope.

## 5. The Giant Magellan Telescope

The GMT consortium is composed of the Carnegie Institution of Washington, Harvard University, Massachusetts Institute of Technology, University of Arizona, University of Michigan, Smithsonian Institution, the University of Texas at Austin and Texas A&M University.

The current GMT design is shown in the figure below. The primary mirror is a set of six disks arranged around a central disk with a secondary hole. The individual mirrors are approximately 8m disks spun-cast at the University of Arizona, with the first off-axis mirror already in production. Although the telescope as a whole is very fast, f/0.7, the individual elements are approximately f/2. The mount has elements of the Large Binocular Telescope concept and the enclosure is a carousel style. The combination is a relatively low risk approach to a telescope with the light gathering power of a 22m and the diffraction limited PSF core of a 25m aperture. The partnership plans to build the telescope either within or near the existing Las Campanas site complex. The current schedule calls for a construction start in 2010 and science operations to begin in 2016. Overall the GMT is taking a relatively conservative approach, designed to minimize both performance and cost risk. At the present time GMT has not announced an expected cost.

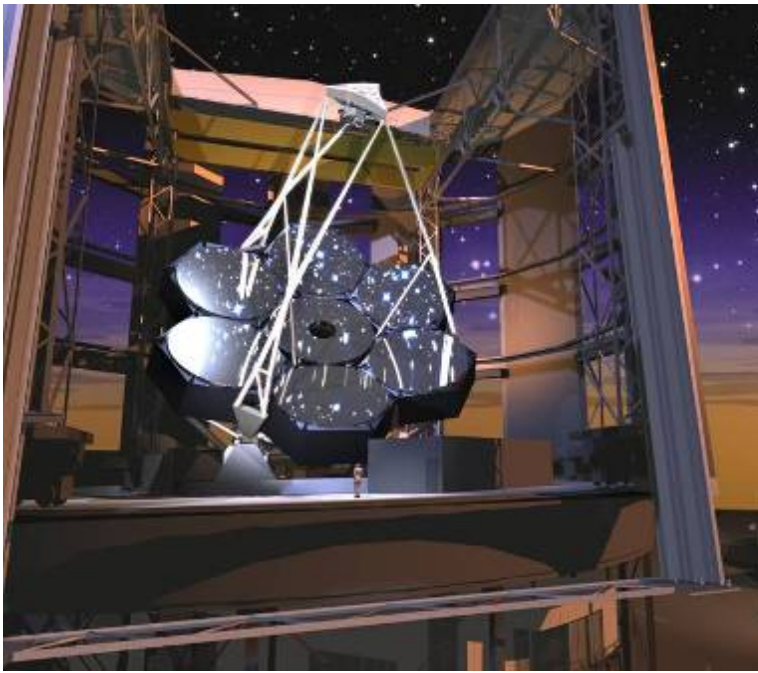

The GMT design. The telescope has 8m Magellan style segments, with relatively large gaps between the segments. The attraction is that the mirrors can be made with a proven technology and the problems of phasing are reduced. The secondary mirror is composed of a set of matched smaller mirrors which simplifies the control problems.

The instrument suite current planned responds to both the scientific interests of the GMT collaboration and the AURA sponsored Giant Segmented Mirror Telescope Science Working group which is responsible for providing a national and international perspective on the scientific requirements of a telescope that US Decadal plan of 2001.

The first generation GMT instruments are summarized in the table below. They cover the full range of science from exo-planets to first light studies. Most of the instruments are capable of operation in natural seeing conditions, which provides a very natural approach to commissioning and then allowing flexible operation. All the instruments sit below the primary mirror in a "stack" that allows all instruments to be available on short notice.

| Instrument | P.I. | Mode | Port |
|---|---|---|---|
| 1. Visible-band Multi-object Spectrograph | S. Shectman | Natural seeing, GLAO | Gregorian |
| 2. High Resolution Visible Spectrograph | P. McQueen | Natural seeing | Folded Port |
| 3. Near-IR Multi-Object Spectrograph | D. Fabricant | Natural Seeing, GLAO | Gregorian |
| 4. Near-IR Extreme AO Imager | L. Close | ExAO | Folded Port |
| 5. Near-IR High Resolution Spectrometers | D. Jaffe | Natural seeing, LTAO | Folded port |
| 6. Mid-IR AO Imager & Spectrograph | P. Hinz | LTAO | Folded port |

The match between the instruments and GSMT SWG science requirements is as follows.

| # | Science Area | Sub-Area | Instruments | Notes |
|---|---|---|---|---|
| 1 | Exoplanets | Direct imaging | 6, 4 | ExAO, Nulling |
|   |   | Disks scattering | 6, 4 |   |
|   |   | Disk emission | 4, 6 | mid-IR |
|   |   | Radial vel. surveys | 3, 5 |   |
| 2 | Solar System | KBOs | 1, 6 |   |
|   |   | Comets & Moons | 3, 5, 4 |   |
| 3 | Star Formation | Embedded clusters | 6, 4 |   |
|   |   | Proper motions | 1 | GLAO critical |
|   |   | Crowded fields | 6, 2, 1 |   |
|   |   | Mass ratios | 6, 3, 5 |   |
| 4 | Stellar Pops | Stellar abundances | 3, 5 |   |
|   |   | Pop. Studies | 1, 6, 3 |   |
| 5 | Black Holes | AGN Environments | 1, 6, 2 | IFU,TF Modes |
|   |   | Velocity structures | 6, 1 | IFU Mode |
| 6 | Dark Energy | SNe monitoring | 6, 1, 2 |   |
|   |   | SNe physics | 1, 2 | Polarim. mode |
| 7 | Galaxy Ass. | Stellar mass density | 1, 2 |   |
|   |   | Internal dynamics | 2, 6 |   |
| 8 | First Light | IGM studies | 1, 3, 2, 5 |   |
|   |   | First Galaxies | 1, 2, 6 |   |

## 6. The Thirty Meter Telescope

The TMT project has four partners, Caltech, the University of California, ACURA (the Canadian consortium) and AURA (the US national consortium) and represents the union of the CELT, GSMT and VLOT projects that undertake initial feasibility studies. At the present time the TMT project has raised a total of US$63M for the detailed design phase (DDP) work on a 30m telescope. The TMT Board has approved a project with a total construction phase budget of US$700M. The current goal of the project is to have full design and cost review completed in the fall of 2006. The DDP work will provide the basis for proposals to the project sponsors that detail the scientific performance capabilities and the final budget. TMT plans to initiate construction in 2009 with scientific operation in 2015.

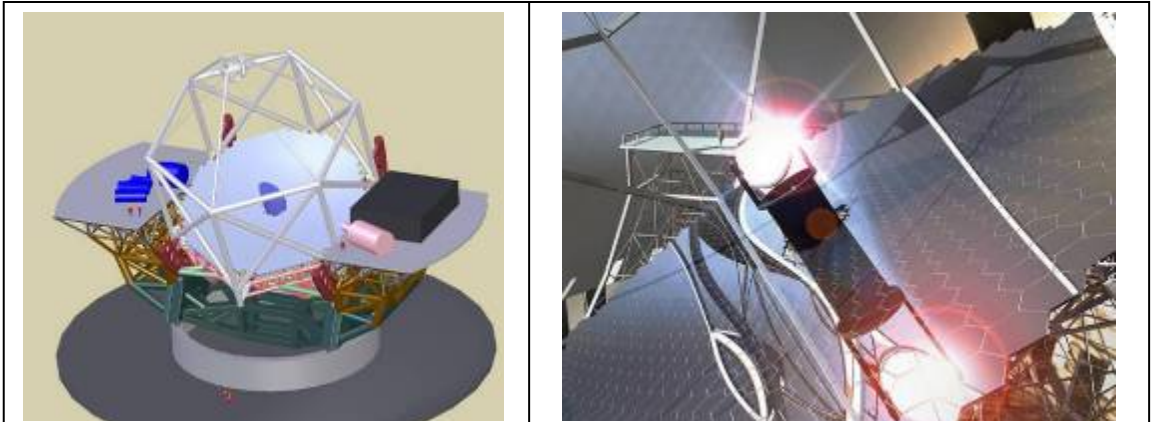

The TMT design. The primary is made of nearly 1m scale segments, as visible in the right hand panel. The large platforms allow several instruments to always be available.

The planned first generation instruments for the TMT necessarily have a great deal of shared conceptual capabilities with those of GMT (or OWL). Given the larger aperture there is more emphasis on providing diffraction limited capabilities from the outset, with an emphasis on the near-infrared part of the spectrum, where the gain from seeing limited is very substantial.

| Instrument | Spectral Resolution | Example Science Cases |
|---|---|---|
| Near-IR DL Spectrometer & Imager (IRIS) | ≤4000 | Assembly of galaxies at large redshift<br>Black holes/AGN/Galactic Center<br>Resolved stellar populations in crowded fields |
| Wide-field Optical Spectrometer (WFOS) | 300 - 5000 | IGM structure and composition 2<z<6<br>High-quality spectra of z>1.5 galaxies suitable for measuring stellar pops, chemistry, energetics |
| Multi-IFU, near-DL, near-IR Spectrometer (IRMOS) | 2000 - 10000 | Near-IR spectroscopic diagnostics of the faintest objects<br>JWST followup<br>Galaxy assembly, chemistry, kinematics during the "epoch of Galaxy Formation" |
| Mid-IR Echelle Spectrometer & Imager | 5000 - 100000 | Physical structure and kinematics of protostellar envelopes<br>Physical diagnostics of circumstellar/protoplanetary disks: where and when planets form during the accretion phase |
| ExAO I (PFI) | 50 - 300 | Direct detection and spectroscopic characterization of extra-solar planets |
| Optical Echelle (HROS) | 30000 - 50000 | Stellar abundance studies throughout the Local Group<br>ISM abundances/kinematics<br>IGM characterization to z~6 |
| MCAO imager (WIRC) | 5 - 100 | Galactic center astrometry; general precision astrometry<br>Stellar populations to 10Mpc |
| Near-IR, DL Echelle (NIRES) | 5000 - 30000 | Precision radial velocities of M-stars and detection of low-mass planets<br>IGM characterizations for z>5.5 |

The TMT is undertaking an extensive campaign of site testing, making cross-calibrated DIMM, MASS, PWV and SODAR measurements in Hawaii, Mexico and Chile. There is an ongoing series of meetings that exchange methods between GMT, TMT and ESO. It is expected that several sites will be found that will be suitable as locations for TMT.

## 7. Japanese Extremely Large Telescope Project

The Japanese have recently released a decadal plan in which having access to a 30m class telescope in 2015 is one of their primary goals. At the present time they are not associated with any particular group. Funded work in Japan is concentrating on novel mirror technologies.

## 8. The Euro-50 Project

The Euro 50 project, growing out of Lund Observatory in Sweden, collaborating with Galway, IAC in Spain, Turku, NPL and UCL in the UK, deserves special note. It is the proto-type effort which started serious thinking about ELTs in the early 1990's and with the addition of a credible AO system helped convince the large telescope community that ELTs were feasible and affordable telescopes. Many of the Euro50 top-level ideas about the primary mirror parameters, optical layout considerations, mechanical structure and have had a strong influence on the feasibility studies of other projects.

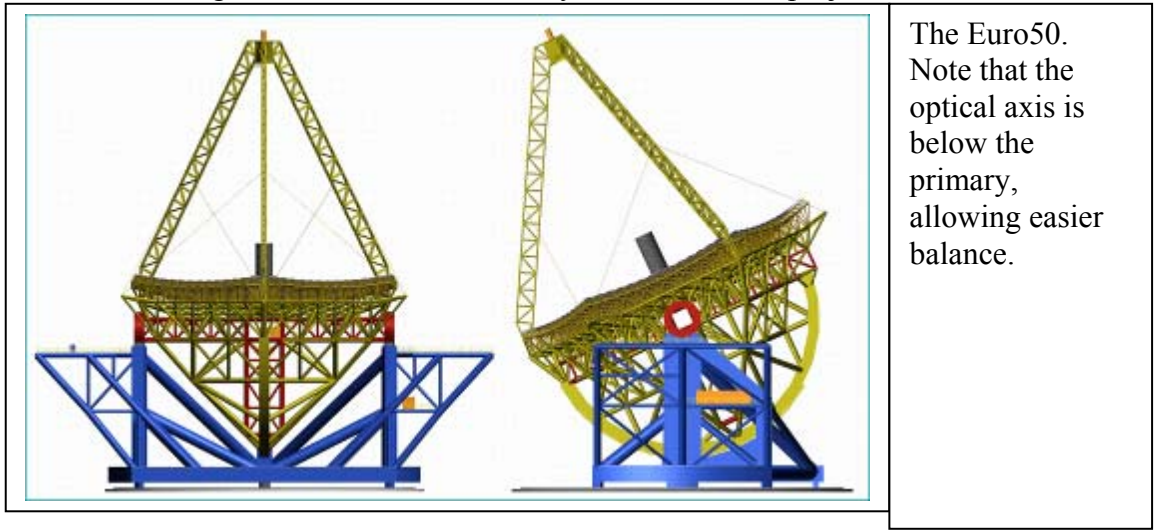

The Euro50. Note that the optical axis is below the primary, allowing easier balance.

## 9. The Overwhelmingly Large Telescope Project

The OWL project entered this meeting with a 100m design concept. Toward the end of the meeting Guy Monnet announced on behalf of ESO that the OWL project intended to concentrate on a revised design in the 40-60m range, with his personal inclination favouring a low-40m aperture target. The ability to make decisions of this magnitude is one of the hallmarks of a successful project office. Moreover it now appears that the community at large has agreed that the 100m scale telescope will be left for the future. This is an extremely important development.

Given the undoubted rewards of an extremely large aperture the OWL project feasibility study elected to make significant design concessions in order to consider whether a 100m aperture could be brought within a plausible cost range. The primary choice to reduce costs is to build a spherical segmented primary mirror. Each hexagonal element is identical and can be mass produced with substantial economies of scale, approximately a factor of ten relative to the single-unit cost. In common with other projects OWL adopted as much off-the-shelf technology as possible, and turns to industry for as many of its needs as possible to inject competition into pricing and to bring the considerable expertise and interest in cutting-edge projects into the OWL project.

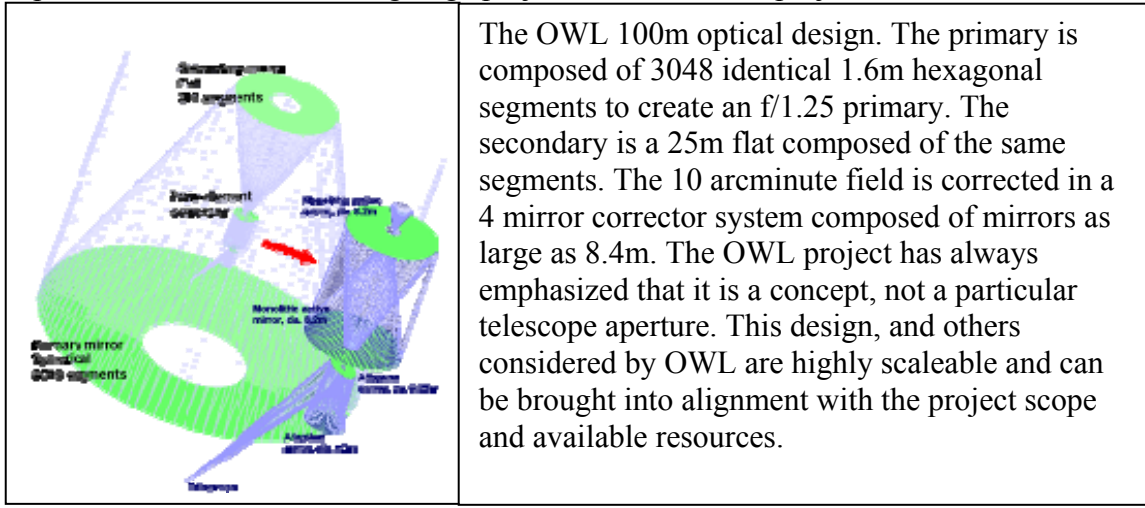

The OWL 100m optical design. The primary is composed of 3048 identical 1.6m hexagonal segments to create an f/1.25 primary. The secondary is a 25m flat composed of the same segments. The 10 arcminute field is corrected in a 4 mirror corrector system composed of mirrors as large as 8.4m. The OWL project has always emphasized that it is a concept, not a particular telescope aperture. This design, and others considered by OWL are highly scaleable and can be brought into alignment with the project scope and available resources.

The 100m design had an estimated cost of $1250MEuro, although it was unclear what further contributions from national observatories would contribute to the total (300 FTE were mentioned in the OWL overview). On Friday Guy Monnet announced that after an external review the OWL project was planning to reduce the budget to approximately 650MEuro. In considering that figure one must recall that much of the ESO work is done in national observatories as contribution to the project. Given that labour dominates the cost the contribution of the size and depth of the considerable talent in the network of European observatories is potentially a huge contribution.

## 10. Concluding remarks

There is an overwhelming science case to build ELTs. All the leading astronomical institutions and nations recognize that these will be such central and powerful facilities that it would be very costly to be left without access to such a facility, both for its own power and the tremendous steering effect it will have on other forefront facilities such as JWST and ALMA which will be completed in about 2013, slightly before the first ELTs come online.

In this situation it would be ideal if more than a single telescope emerged, ideally with some degree of complementarity to maximize observing opportunities and minimize the number of extremely expensive large (8m telescope size class) instruments that are "dark" when another instrument is in use.

The Cape Town ELT meeting provided an interesting snapshot of the state of several telescope projects. The participants saw one dramatic development over the course of the meeting. It seems likely that there will several other significant developments before the first one of these comes online, after which at least one more is likely to emerge fairly quickly.